\documentclass[%
 aip,
 amsmath,amssymb,
 reprint,%
 floatfix,
]{revtex4-1}

\usepackage{setspace}
\usepackage{relsize}
\usepackage{verbatim}
\usepackage{graphicx}
\usepackage{dcolumn}
\usepackage{bm}
\usepackage{adjustbox}
\usepackage[utf8]{inputenc}
\usepackage[T1]{fontenc}
\usepackage{anyfontsize}
\usepackage{mathptmx}
\usepackage{etoolbox}
\usepackage{xcolor}
\usepackage{comment}
\usepackage{setspace}

\makeatletter
\def\@email#1#2{%
 \endgroup
 \patchcmd{\titleblock@produce}
  {\frontmatter@RRAPformat}
  {\frontmatter@RRAPformat{\produce@RRAP{*#1\href{mailto:#2}{#2}}}\frontmatter@RRAPformat}
  {}{}
}%
\makeatother
\begin{document}

\preprint{AIP/123-QED}

\title{Mitigation of limit cycle oscillations in a turbulent thermoacoustic system \\ via delayed acoustic self-feedback}
\author{Ankit Sahay}
\email{ankitsahay02@gmail.com}
 
\author{Abhishek Kushwaha}%
\author{Samadhan A. Pawar}
\author{Midhun P. R.}
\author{Jayesh M. Dhadphale}
\author{R. I. Sujith}
\affiliation{Department of Aerospace Engineering, Indian Institute of Technology Madras, Chennai, Tamil Nadu 600036, India.}
\date{\today}

\begin{abstract}
We report the occurrence of amplitude death (AD) of limit cycle oscillations in a bluff body stabilized turbulent combustor through delayed acoustic self-feedback. Such feedback control is achieved by coupling the acoustic field of the combustor to itself through a single coupling tube attached near the anti-node position of the acoustic standing wave. We observe that the amplitude and dominant frequency of the limit cycle oscillations gradually decrease as the length of the coupling tube is increased. Complete suppression (AD) of these oscillations is observed when the length of the coupling tube is nearly $3/8$ times the wavelength of the fundamental acoustic mode of the combustor. Meanwhile, as we approach this state of amplitude death, the dynamical behavior of acoustic pressure changes from the state of limit cycle oscillations to low-amplitude chaotic oscillations via intermittency. We also study the change in the nature of the coupling between the unsteady flame dynamics and the acoustic field as the length of the coupling tube is increased. We find that the temporal synchrony between these oscillations changes from the state of synchronized periodicity to desynchronized aperiodicity through intermittent synchronization. Furthermore, we reveal that the application of delayed acoustic self-feedback with optimum feedback parameters completely disrupts the positive feedback loop between hydrodynamic, acoustic, and heat release rate fluctuations present in the combustor during thermoacoustic instability, thus mitigating instability. We anticipate this method to be a viable and cost-effective option to mitigate thermoacoustic oscillations in turbulent combustion systems used in practical propulsion and power systems.
\end{abstract}

\maketitle

\begin{quotation}

Thermoacoustic instabilities are large amplitude, self-sustained periodic oscillations observed in practical gas turbines and rocket engines used for propulsion and power generation applications. Such instability results from positive feedback between the heat release rate and acoustic pressure fluctuations. The presence of these large amplitude acoustic pressure oscillations can severely affect the structural integrity of a combustor. Over the years, various active and passive control strategies have been developed to mitigate thermoacoustic instability. Recently, time delayed feedback involving the use of a connecting tube has been shown to suppress limit cycle oscillations in different laminar systems such as acoustic pipelines and Rijke tubes. However, unlike laminar thermoacoustic systems, a turbulent thermoacoustic system is a complex system where the local closed-loop interaction between hydrodynamic, acoustic, and heat release rate fluctuations leads to an emergence of periodic (ordered) oscillations during the onset of thermoacoustic instability. In the present study, we show the use of time delayed feedback to suppress limit cycle oscillations in a single bluff body stabilized turbulent combustor. We couple (feedback) the acoustic field of the combustor to itself through a connecting tube. The length of the connecting tube is proportional to the time delay in the feedback; therefore, we refer to this method as delayed acoustic self-feedback in the present study. We show that at the appropriate length and diameter of the coupling tube, the proposed method completely disrupts the complex interactions inside the reaction field of the turbulent combustor during thermoacoustic instability, leading to the suppression of amplitude of limit cycle oscillations by more than 90\%. During the state of suppression, the amplitude of acoustic pressure fluctuations is comparable to that observed for the state of stable operation (i.e., combustion noise) of the combustor.
\end{quotation}

\section{\label{sec:Introduction}Introduction}

Thermoacoustic instabilities have proven to be a major impediment to the development of low-emission gas turbine engines used for propulsion and power generation applications \cite{lieuwen2005combustion}. Such instabilities lead to ruinously large amplitude pressure oscillations that are established when positive feedback develops between the acoustic field and the heat release rate fluctuations in the reaction field of the combustor \cite{sujith2021thermoacoustic}. The presence of these instabilities results in serious performance losses, structural damages, and reduced operational range \cite{culick2006unsteady}. Therefore, it is necessary to find ways to mitigate thermoacoustic instability in the course of developing new dynamically stable combustion systems.

Traditionally, different mechanisms of closed-loop and open-loop active controls have been developed for suppressing thermoacoustic instability \cite{dowling2005feedback}. In these methods, the dynamics of a thermoacoustic system is forced using external perturbations of specific amplitudes and frequencies. However, these methods suffer from several limitations, such as the use of complex electromechanical components, lack of reliability of sensors while working in the hostile conditions of practical combustors, and high maintenance and replacement costs. Another way to mitigate thermoacoustic instability is to use passive damping devices such as perforated liners, quarter and half-wave resonators, Helmholtz resonators \cite{zhao2015review}, Herschel-Quincke tubes \cite{rajaram2012attenuation}, modification of the fuel injector geometry/location \cite{steele2000passive, smith2005combustion}, and secondary air/fuel injections \cite{schadow1992combustion, huang2009dynamics}. Engine manufacturers have used such passive damping devices to suppress thermoacoustic instability in practical combustors \cite{lieuwen2005combustion, bellucci2004use}. 

Recently, a concept from synchronization theory \cite{lakshmanan1996chaos, pikovsky2003synchronization}, called mutual coupling of oscillators, has been adopted to suppress limit cycle oscillations (LCOs) in two or more thermoacoustic systems \cite{sujith2021thermoacoustic}. At appropriate values of coupling parameters, the coupled systems approach the same steady state of oscillation quenching, known as amplitude death \cite{zou2021quenching}. Through rigorous experimental and numerical analysis, amplitude death \cite{thomas2018effect1, hyodo2020suppression, ghosh2022occasional} and partial amplitude death \cite{dange2019oscillation, srikanth2021dynamical} have been studied in coupled laminar thermoacoustic systems \cite{doranehgard2022quenching}. Here, partial amplitude death is a phenomenon in which some of the oscillators are in the suppressed state, but the others continue their oscillations \cite{atay2003total}. In addition, a few studies have investigated the dynamics of coupled turbulent combustors \cite{thomas2018effect, jegal2019mutual, moon2020mutual, guan2021low, pedergnana2022coupling, fournier2021low, moon2020cross, pedergnana2022steady}, and have reported the presence of amplitude death \cite{jegal2019mutual}. Here, the mutual coupling between the systems is achieved by using one (or more than one) connecting tube of a fixed length and diameter. In all the studies, it is often assumed that an increase in the length of the coupling tube correspondingly increases the delay time in the coupling of the acoustic fields in the systems \cite{dange2019oscillation, sahay2021dynamics}.

Until now, the concept of amplitude death has been proposed to mitigate thermoacoustic instability in multiple coupled thermoacoustic systems. A majority of these studies have extensively focused on suppressing thermoacoustic instability in Rijke tubes. Rijke tubes are mostly laminar thermoacoustic systems and are relatively simpler when compared to practical turbulent thermoacoustic systems \cite{manoj2022rijke}.  In a Rijke tube, the system behavior loses the stability of the steady state and starts exhibiting LCOs via a Hopf bifurcation due to a change in the control parameter \cite{manoj2022rijke}. In contrast, in a typical turbulent combustion system, the system behavior transitions from high-dimensional chaos to LCOs (thermoacoustic instability) via intermittency \cite{nair2014intermittency}. Intermittency is characterized by epochs of low-amplitude chaotic oscillations interspersed with seemingly random bursts of high-amplitude periodic oscillations. Due to the turbulent flow field, interactions between hydrodynamic, acoustic, and heat release rate fluctuations are significantly complex when compared to laminar systems \cite{sujith2021thermoacoustic}. The nonlinear interplay of such different subsystems at local scales can lead to the emergence of collective behavior at the global scale. The process of larger entities, patterns, and regularities emerging through interactions between constituent entities that are unable to exhibit these features on their own is known as "emergence," and it is a key component of complex systems. Large coherent vortices and an ordered acoustic field develop from the background of turbulent combustion during the onset of thermoacoustic instability in a turbulent combustion system.

In this paper, we demonstrate the effectiveness of coupling the acoustic field of a single turbulent combustor with itself to break the closed-loop interaction between the heat release rate and the acoustic pressure fluctuations and, thus, mitigating thermoacoustic instability. This method of coupling the acoustic field with itself after a particular duration is referred to as delayed acoustic self-feedback \cite{biwa2016suppression, srikanth2021selfcoupling}. As compared to the traditional closed-loop active controls \cite{dowling2005feedback} where the acoustic field in the combustor is subjected to its own feedback after processing it via digital delay lines and power amplifiers to alter the inlet conditions of the combustor \cite{annaswamy2002active}, the method of self-feedback is realized by continuously making the acoustic field in the system interact with itself through a connecting tube. The connecting tube induces a time delay in the feedback, where the time delay is proportional to the length of the tube \cite{biwa2015amplitude, dange2019oscillation}. Time delayed feedback introduced electronically has been successfully used in diverse experimental systems, such as lasers \cite{basso1998optimal}, gas-discharge \cite{pierre1996controlling}, hydrodynamic \cite{luthje2001control}, electrochemical \cite{parmananda1999stabilization}, and ferromagnetic \cite{benner2002control} systems. Recently, time delayed feedback through the use of a connecting tube has also been used to suppress LCOs in the acoustic field of different systems, such as electroacoustic system \cite{biwa2016suppression}, acoustic pipeline \cite{lato2019passive}, and horizontal Rijke tube \cite{srikanth2021selfcoupling}, that does not involve turbulent flow. These studies have shown that subjecting a system to delayed acoustic self-feedback affects its bifurcation characteristics \cite{srikanth2021selfcoupling}, and the suppression of LCOs is realized only when a tube of a length close to the odd multiple of the half-wavelength of the anticipated acoustic standing wave is used \cite{biwa2016suppression, lato2019passive, srikanth2021selfcoupling}. We here discuss the application of this method to suppress LCOs in a single turbulent combustion system. Toward this purpose, we systematically address the following questions in this paper - (i) how effective is delayed acoustic self-feedback in disrupting the complex interactions between the acoustic pressure and the heat release rate fluctuations and consequently mitigating thermoacoustic instability in a turbulent combustor? (ii) What is the nature of the transition of acoustic pressure fluctuations during the suppression of thermoacoustic instability? Moreover, (iii) how do the temporal and spatiotemporal couplings affect the heat release rate and acoustic pressure oscillations due to delayed acoustic self-feedback?

\begin{figure*}[t!]
\includegraphics[width=0.6\textwidth]{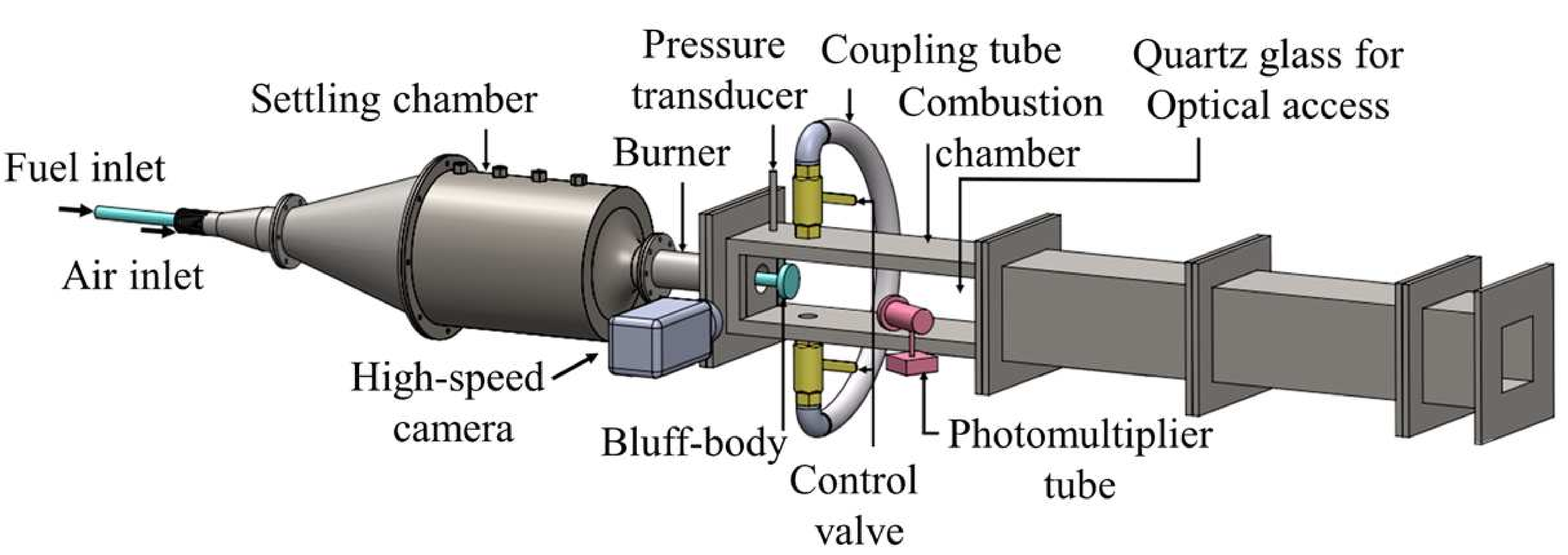}
\caption{\label{TARA_fig} The schematic of a turbulent bluff body stabilized combustor subjected to delayed acoustic self-feedback using a single coupling tube.}
\end{figure*}

We conduct experiments on a bluff body stabilized turbulent combustor to answer the aforementioned questions. We report the mitigation of thermoacoustic instability in the combustor through delayed acoustic self-feedback when the connecting tube is attached at the acoustic pressure anti-node of the combustor. Previous studies have demonstrated that the optimal location for pressure actuation is near the pressure maxima of the acoustic standing wave \cite{magri2013sensitivity, srikanth2021selfcoupling, skene2022phase}. The amplitude of acoustic pressure fluctuations in the state of amplitude death is comparable to that observed during the stable operation of the combustor. We find that the suppression of thermoacoustic instability depends on various parameters, i.e., the length and diameter of the coupling tube and the amplitude of thermoacoustic instability observed prior to the initiation of coupling. As we approach the state of amplitude death, acoustic pressure fluctuations change their behavior from LCOs to low- amplitude chaotic oscillations via intermittency. Also, the coupled behavior of acoustic pressure and heat release rate fluctuations transitions from the state of phase synchronization to desynchronization via intermittent synchronization. Furthermore, the coherent regions of acoustic power production observed in the spatial field of the combustor during the state of thermoacoustic instabilities disintegrate completely during the state of amplitude death.

The rest of the paper is outlined as follows. In Sec.~\ref{sec:model_Experimental_Setup}, we describe the experimental setup of a bluff body stabilized turbulent combustor used in this study. In Sec.~\ref{sec:Res_Disc_A} - \ref{sec:Res_Disc_B}, we perform time series analyses of global heat release rate and acoustic pressure fluctuations for different conditions of delayed acoustic self-feedback. In Sec.~\ref{sec:Res_Disc_C}, we compare the spatiotemporal changes in the combustor as limit cycle oscillations are suppressed. Finally, we summarize the key findings from the study in Sec.~\ref{sec:Conclusion}.

\section{\label{sec:model_Experimental_Setup}Experimental Setup}

We experimentally demonstrate the application of delayed acoustic self-feedback of the acoustic field to mitigate thermoacoustic instability in a turbulent bluff body stabilized dump combustor (Fig.~\ref{TARA_fig}). The combustor has a cross section of $90 \times 90$ $\text{mm}^2$ and a length of $L_{\text{duct}} = 1060$ $\text{mm}$. Air at ambient conditions first enters the settling chamber, which ensures that the flow entering the combustor is isolated from the upstream disturbances. The fuel (liquefied petroleum gas - 60\% butane + 40\% propane) is injected just upstream of the dump plane (approx. 59 mm upstream of the bluff body) through four holes of 1 mm diameter drilled symmetrically around the circular shaft that holds the bluff body in the combustor. A disk-shaped bluff body of thickness 10 mm and diameter 47 mm is used as the flame stabilizer and is positioned 30 mm downstream of the dump plane. The combustion products are exhausted through a long duct into the atmosphere via a decoupler. The decoupler is a large chamber used to keep the combustor exit at an acoustically open boundary state and to decrease acoustic losses to the environment brought on by acoustic radiation.

Digital mass flow controllers (Alicat Scientific, MCR Series) are used to control the flow rates of both air and fuel separately. During the experiments, we fix the fuel flow rate at a particular value and increase the air flow rate until thermoacoustic instability is established in the system. The increase in the air flow rate causes a decrease in the global equivalence ratio. The global equivalence ratio is calculated as $\phi = (\dot{m_f}/\dot{m_a})_{\text{actual}}/(\dot{m_f}/\dot{m_a})_{\text{stoichiometry}}$. The global equivalence ratio is varied from 1.00 to approximately 0.6, with an uncertainty of $\pm 0.02$. The air flow rate varies quasi-statically from 527 to 900 SLPM (standard liter per minute), with a maximum uncertainty in the flow rate of air being $ \pm 10.24$ SLPM. The fuel flow rate is kept constant at 34 SLPM, and the maximum uncertainty in the fuel flow rate is $ \pm 0.44$ SLPM. The Reynolds number of the air flow increases in the range of 14,500 to 25,000, with a maximum uncertainty of $\pm 400$. The Reynolds number was computed using the expression $Re = 4 \dot{m}D_1 / \pi \mu D^2_0$, where $\dot{m}$ is the mass flow rate of the fuel-air mixture, $D_0$ is the diameter of the burner, $D_1$ is the diameter of the circular bluff body, and $\mu$ is the dynamic viscosity of the fuel-air mixture at standard operating conditions. We use the dynamic viscosity expression given by Wilke \cite{wilke1950viscosity} to calculate the Reynolds number for a binary gas mixture of air and fuel. The air-fuel mixture is ignited at the dump plane using a spark plug connected to an 11 kV ignition transformer.

%
%

Delayed acoustic self-feedback is established in the combustor using a single coupling tube made of a flexible stainless steel braided hose of length $L_{\text{c}}$ and internal diameter $d_{\text{c}}$ (refer to Fig.~\ref{TARA_fig}). The value of $L_{\text{c}}$ is varied from 1000 to 2000 mm in steps of 100 mm, while the value of $d_{\text{c}}$ is varied from 6.35 to 25.4 mm ($d_{\text{c}} = 6.35$, $9.525$, $12.7$, $19.05$, and $25.4$ mm). The coupling tube is attached at an axial distance of 70 mm from the dump plane on two opposite faces of the combustor walls. This position is near the anti-node of the acoustic pressure standing wave that, in turn, induces stronger acoustic feedback in the coupled system. As mentioned in Sec. \ref{sec:Introduction}, previous studies have shown that the optimal location for pressure actuation is near the pressure maxima of the acoustic standing wave \cite{magri2013sensitivity, srikanth2021selfcoupling, skene2022phase}. We have not used any other position on the combustor duct to attach the coupling tube. Ball-type valves are manually operated to switch on and off the delayed self-feedback of the acoustic field in the system. We first establish thermoacoustic instability in the system and then switch on the delayed feedback by opening the valves of a coupling tube of specific length $L_{\text{c}}$ and internal diameter $d_{\text{c}}$. 

\begin{figure*}
\includegraphics[width=1\textwidth]{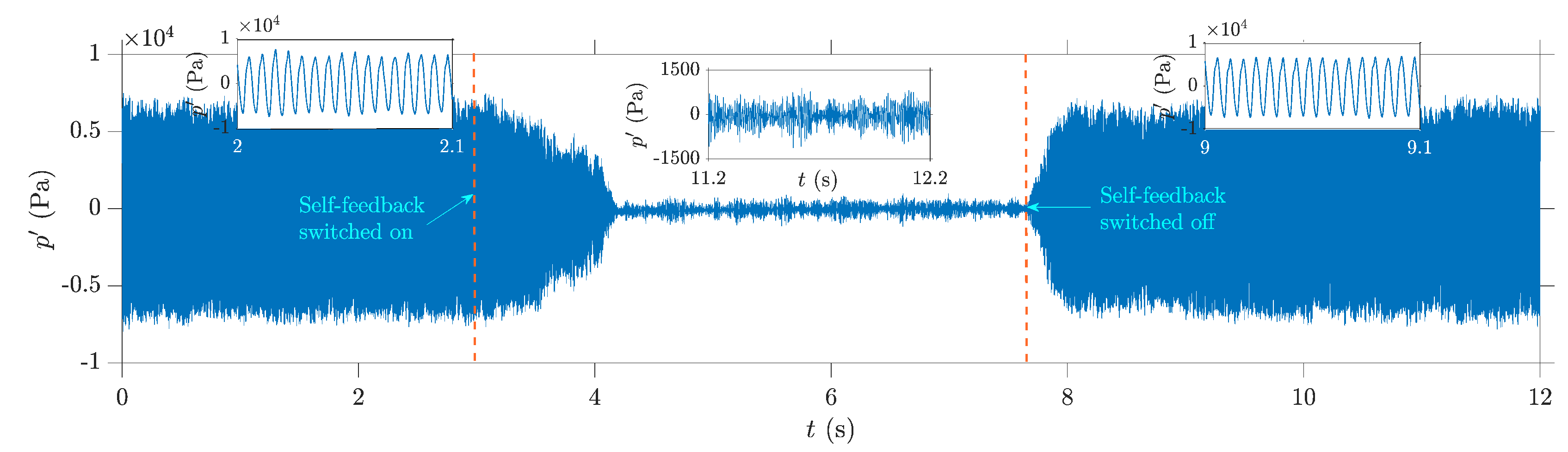}
\caption{\label{fig2a} A representative time series of acoustic pressure $p^\prime$ illustrating the effect of delayed acoustic self-feedback on the limit cycle oscillations. The enlarged portions show the dynamics of $p^\prime$ during different stages of delayed acoustic self-feedback. The dimensions of the coupling tube are $L_{\text{c}} = 1400$ mm and $d_{\text{c}} = 25.4$ mm.}
\end{figure*}

The acoustic pressure fluctuations $p^{\prime}$ are measured using a PCB103B02 piezoelectric transducer (sensitivity: 217.5 mV/kPa and uncertainty: $\pm$0.15 Pa) mounted on the combustor wall at 20 mm from the dump plane. A photomultiplier tube (PMT, Hamamatsu H10722-01) equipped with a CH* filter (wavelength of 430 nm and 12 nm FWHM) is used to capture the global heat release rate fluctuations in the flame $\dot{q}^{\prime}$. We simultaneously acquired both $p^{\prime}$ and global $\dot{q}^{\prime}$ signals for 3 s  at a sampling rate of 10 kHz using an A/D card (NI-6143, 16 bit). The frequency bin size in the power spectrum is 0.3 Hz. High-speed CH* chemiluminescence images of the flame are simultaneously captured with $p^{\prime}$ and $\dot{q}^{\prime}$ signals at 2000 Hz for 3 s using a CMOS camera (Phantom - V12.1 with a ZEISS 50 mm camera lens). The image of the flow field from the dump plane spans 90 $\times$ 120 $\text{mm}^2$ with a resolution of 574 $\times$ 764 pixels. To ensure consistency in experimental conditions for different trials, the experiments were performed only when the decay rate of periodic perturbations in the acoustic pressure generated through loudspeakers, measured in the absence of flow, remained constant at $11 \pm 1.5$ s$^{-1}$. This decay rate is ensured in the combustor irrespective of the presence of the coupling tube.

\section{\label{sec:Results_and_Discussion}Results and Discussion}

In this section, we characterize the effects of delayed acoustic self-feedback on the suppression of thermoacoustic instability in the turbulent combustor. We first discuss the dynamical behavior of the acoustic pressure signal ($p^\prime$) alone and then present the change in the coupled behavior of both global heat release rate ($\dot{q^\prime}$) and acoustic pressure ($p^\prime$) fluctuations during the suppression of thermoacoustic instability.

\subsection{\label{sec:Res_Disc_A}Route from thermoacoustic instability to the state of suppression}

In Fig.~\ref{fig2a}, we show a representative time series of the acoustic pressure fluctuations under the influence of an optimal delayed acoustic self-feedback induced using a coupling tube of length $L_{\text{c}} = 1400$ mm and an internal diameter of $d_{\text{c}} = 25.4$ mm. Prior to the initiation of feedback, we observe LCOs with a root-mean-square (RMS) value of $p^\prime_{0,\text{rms}}=4390$ Pa and a frequency of $164$ Hz in the combustor. The subscript $0$ in $p^\prime_{0,\text{rms}}$ indicates the state of thermoacoustic instability in the absence of delayed acoustic self-feedback. The application of delayed acoustic self-feedback causes a maximum reduction in the RMS value of the LCOs to about $p^\prime_{\text{rms}}=230$ Pa, where the acoustic pressure fluctuations exhibit chaotic oscillations. We refer to this state of complete suppression of thermoacoustic instability in a turbulent combustor as amplitude death \cite{reddy2000dynamics}. We emphasize that the turbulent combustor is not perfectly silent during its stable state of operation, where we notice  $p^\prime_{0,\text{rms}} = 120 \pm 50$ Pa. Hence, we do not have a near 100\% reduction in the RMS value of acoustic pressure fluctuations during the state of amplitude death as that observed in laminar thermoacoustic systems \cite{srikanth2021selfcoupling}. The removal of delayed acoustic self-feedback at around $t = 7.6$ s revives the LCOs, and the combustor returns to the original state of thermoacoustic instability.

\begin{figure}
\includegraphics[width=0.45\textwidth]{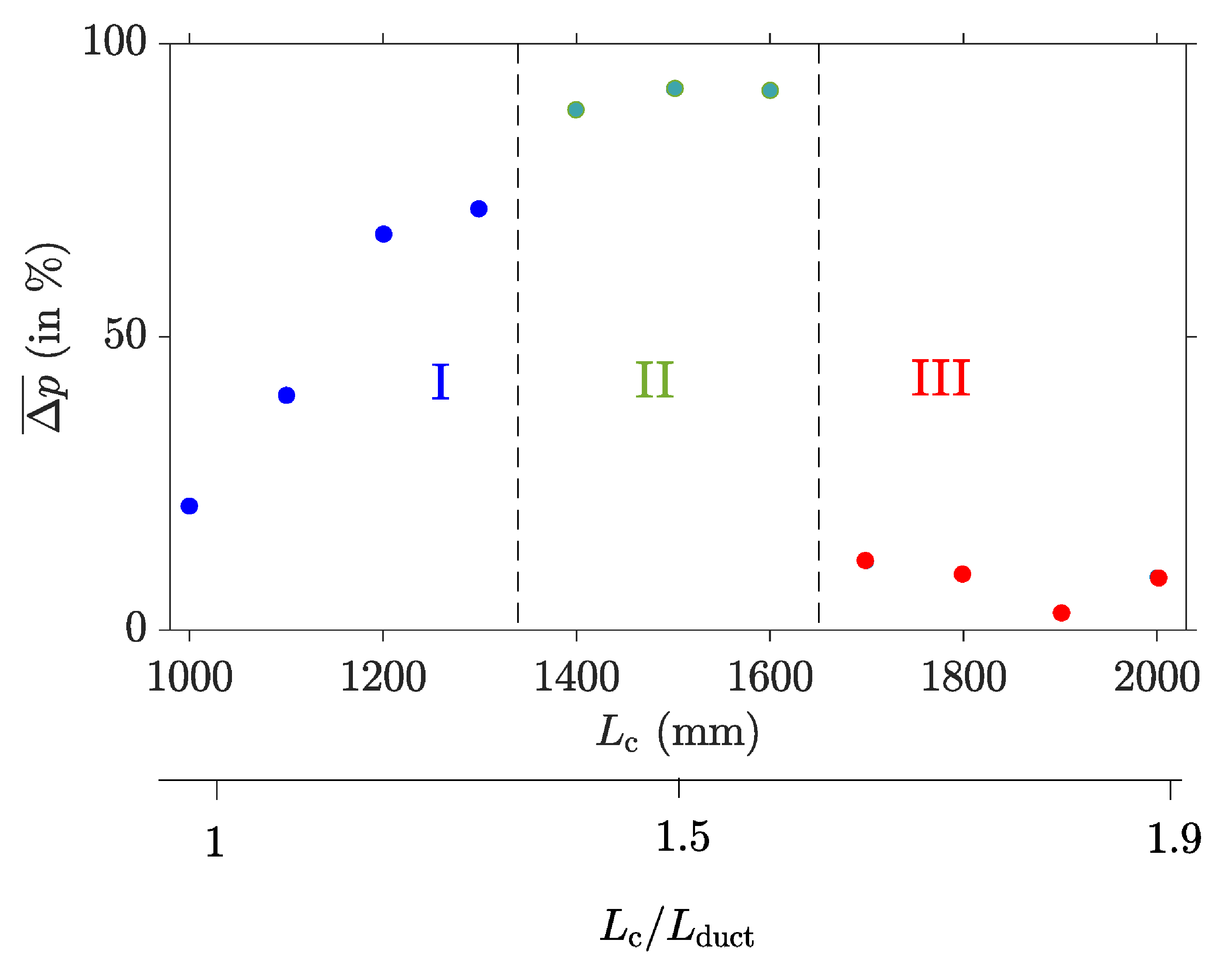}
\caption{\label{fig2} Effect of delayed acoustic self-feedback, induced by increasing the length of the coupling tube $L_{\text{c}}$, on the suppression of acoustic pressure fluctuations indicated by the normalized suppression percentage $\overline{\Delta {p}}$. The regions marked in I, II, and III denote the partial suppression, complete suppression (amplitude death), and no suppression  of LCOs ($p^\prime_{0,\text{rms}} \approx 3200$ Pa), respectively. The value of $d_{\text{c}}$ is fixed at 25.4 mm.}
\end{figure}

In Fig.~\ref{fig2}, we show the percentage change in the RMS value of $p^{\prime}$ signals, i.e., $\overline{\Delta {p}}$, as a function of the length of the coupling tube ($L_{\text{c}}$) when its internal diameter is kept constant at 25.4 mm. Here, $\overline{\Delta {p}}$ indicates the normalized difference between the RMS values of the $p^{\prime}$ signal during the state of thermoacoustic instability when the delayed acoustic self-feedback is off ($p^\prime_{0,\text{rms}}$) and that of $p^{\prime}$ signal when the delayed acoustic self-feedback is switched on, i.e., $\overline{\Delta {p}}= ((p^\prime_{0,\text{rms}}-p^\prime_{\text{rms}})/p^\prime_{0,\text{rms}}) \times 100$. We define the partial suppressed state as the state where the suppression of limit cycle oscillations (normalized suppression percentage indicated by $\overline{\Delta {p}}$) is greater than 20\% but less than 80\% (region I in Fig. \ref{fig2}). The states exhibiting suppression less than 20\% and greater than 80\% are defined as the states with no suppression and complete suppression (amplitude death), respectively. These states are illustrated as regions II and III in Fig. \ref{fig2}, respectively. We notice that as the length of the coupling tube is increased, the amplitude suppression of thermoacoustic instability also correspondingly increases (region I in Fig. \ref{fig2}). Once $L_{\text{c}}$ is above a critical value, we observe amplitude death in the system. If $L_{\text{c}}$ is increased post the region of amplitude death (i.e., in region III of Fig. \ref{fig2}), $\overline{\Delta {p}}$ dips suddenly below 20\% as thermoacoustic instabilities are nearly unaffected by the presence of the coupling tube. This further indicates that there exists a critical length of the coupling tube for a given value of the tube diameter ($d_{\text{c}}$), where amplitude death is observed due to delayed acoustic self-feedback. Also, we note that the variation of $\overline{\Delta {p}}$ is asymmetric across the amplitude death region (in regions I and III in Fig. \ref{fig2}). Furthermore, we find that the optimum value of $L_{\text{c}}$ corresponding to the amplitude death of thermoacoustic instability is around 1.5 times the length of the combustor (i.e., $L_{\text{c}}/L_{\text{duct}} \approx 1.5$). Since the combustor geometry can be modeled as an open-closed acoustic duct \cite{seshadri2016bifurcation}, and as the combustor exhibits the first acoustic mode during the state of thermoacoustic instability (i.e., $L_{\text{duct}} = \lambda/4$), the optimum value of $L_{\text{c}}$ corresponding to the state of amplitude death is approximately equal to $\dfrac{3}{2} \times L_{\text{duct}} = \dfrac{3}{2} \times \dfrac{\lambda}{4} = \dfrac{3 \lambda}{8}$, where $\lambda$ is the wavelength of the acoustic standing wave developed in the combustor duct during the state of thermoacoustic instability.

\begin{figure*}
\includegraphics[width=1\textwidth]{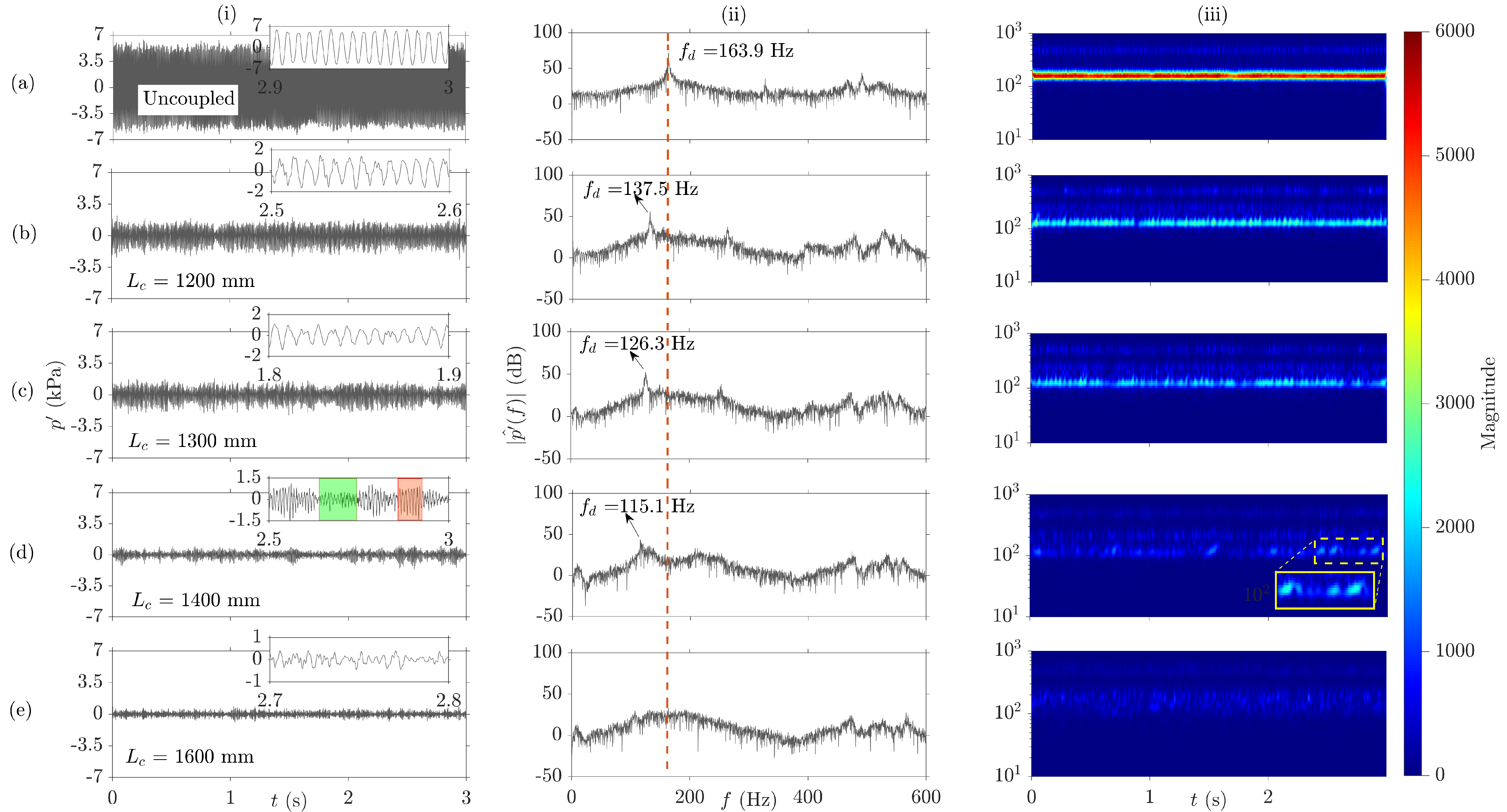}
\caption{\label{fig3} (i) Time series, (ii) power spectrum, and (iii) scalograms of $p^\prime$ as the system behavior transitions from a state of (a) thermoacoustic instability to (e) amplitude death of oscillations via (b-d) intermittency. A vertical dotted red line in (ii) indicates the natural frequency of thermoacoustic instability in the absence of delayed acoustic self-feedback. Zoomed regions of plots are shown in insets, where the green and red shaded regions in (i)-(d) highlight the aperiodic and periodic epochs of $p^\prime$ signal during the state of intermittency, respectively.}
\end{figure*}

Figure~\ref{fig3} shows the change in the characteristics of $p^{\prime}$ signal in the absence of coupling (Fig.~\ref{fig3}a) and when the combustor is subjected to delayed acoustic self-feedback for different lengths of the coupling tube (Figs.~\ref{fig3}b-e). In the absence of such feedback, during thermoacoustic instability (Fig.~\ref{fig3}a), we observe large amplitude LCOs in the $p^{\prime}$ signal (Fig.~\ref{fig3}a-i) with a sharp spectral peak at 163.9 Hz corresponding to the fundamental mode of the combustor (Fig.~\ref{fig3}a-ii). The scalogram of the $p^{\prime}$ signal shows a continuous distribution of the spectral power throughout the time in a narrow frequency band around 163.9 Hz (Fig.~\ref{fig3}a-iii). The introduction of delayed acoustic self-feedback leads to a significant change in the dynamical behavior of the $p^{\prime}$ signal. As the suppression of thermoacoustic instability is approached on increasing $L_{\text{c}}$, the dynamics of $p^{\prime}$ signal changes from the state of large amplitude LCOs (Fig.~\ref{fig3}b) to low-amplitude chaotic oscillations (Fig.~\ref{fig3}e) via a regime of intermittent oscillations (Figs.~\ref{fig3}c and \ref{fig3}d). During the occurrence of intermittent oscillations, epochs of low-amplitude aperiodic fluctuations appear amidst epochs of large amplitude periodic oscillations in an apparently random manner (see inset in Fig.~\ref{fig3}d-i and Fig.~\ref{fig3}d-iii). With an increase in $L_{\text{c}}$, we also notice that the dominant frequency ($f_d$) of $p^{\prime}$ signal shifts gradually towards lower values (compare Fig.~\ref{fig3}a-ii to Fig.~\ref{fig3}e-ii). Similar observation has been observed in a variety of systems subjected to self-delayed feedback \cite{flunkert2007suppressing, scholl2009time, dai2015time}, where  a change of the natural frequency of an oscillator in a sawtooth fashion is reported with an increase in the coupling time delay. As the change in $L_{\text{c}}$ corresponds to a change in the acoustic feedback delay in our system, we attribute the change in the peak spectral frequency ($f_d$) of $p^{\prime}$ signals to the change in the length of the coupling tube. Furthermore, we notice that the spectral magnitude of $f_d$ decreases continuously such that the power spectrum appears broadband during the state of amplitude death. The scalogram plots during this transition show an increase in discontinuities in the variation of the dominant spectral amplitude of the signal (Fig.~\ref{fig3}b-iii to Fig.~\ref{fig3}d-iii), where these discontinuities increase as we approach the state of amplitude death (Fig.~\ref{fig3}e-iii). 

\begin{figure}
\includegraphics[width=0.45\textwidth]{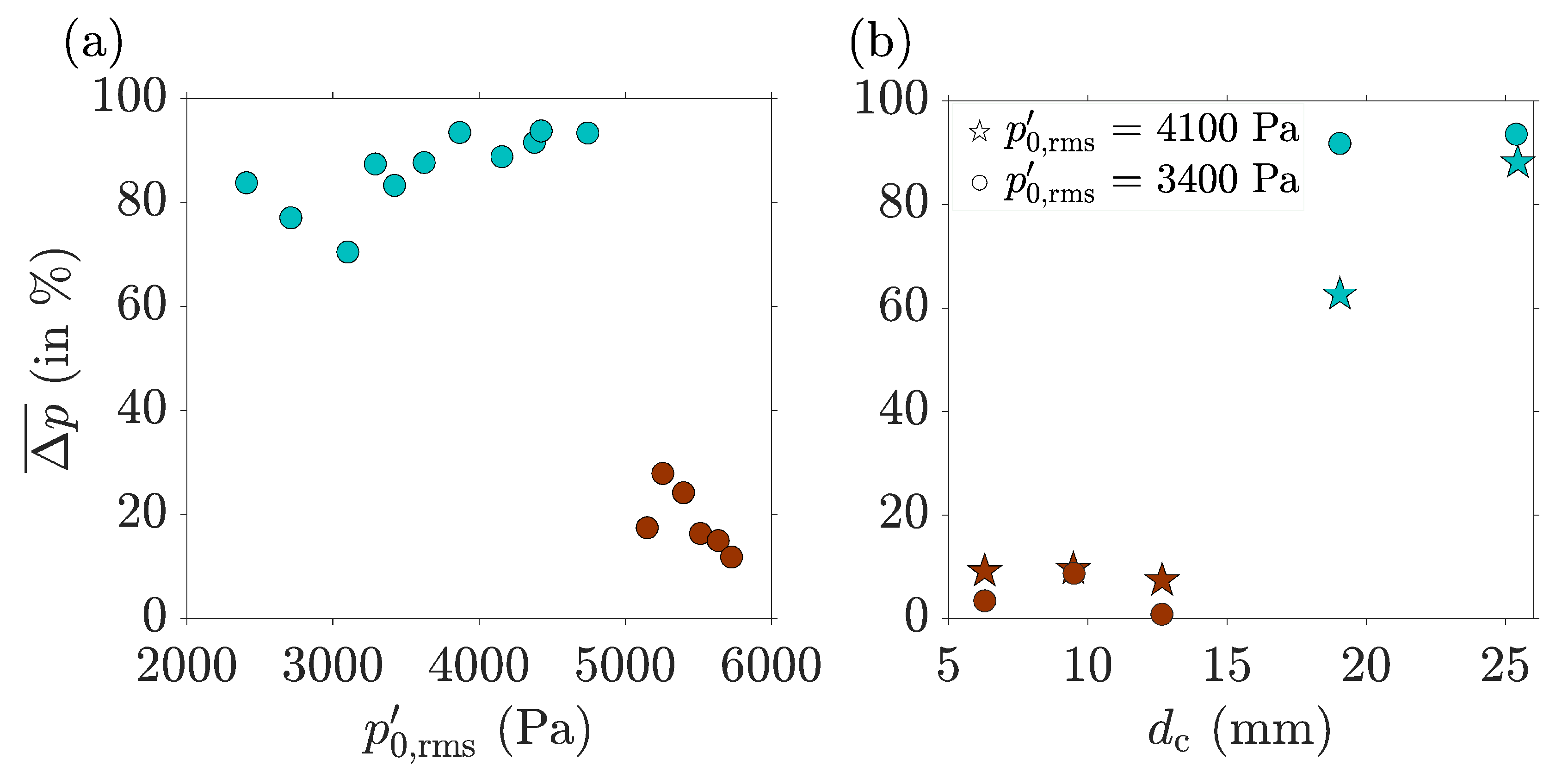}
\caption{\label{diff_amp_dia} The percentage suppression of $p^\prime$ signals ($\overline{\Delta {p}}$) when delayed acoustic self-feedback is applied to (a) thermoacoustic instability of different amplitudes ($p^\prime_{0,\text{rms}}$) for constant values of $L_{\text{c}} = 1400$ mm and $d_{\text{c}} = 25.4$ mm, and (b) thermoacoustic instability of two amplitudes ($p^\prime_{0,\text{rms}} \approx 3400$ Pa and $4100$ Pa) for different coupling tube diameters ($d_{\text{c}} = 6.35$, $9.525$, $12.7$, $19.05$, and $25.4$ mm) at a fixed value of $L_{\text{c}} = 1400$ mm.}
\end{figure}

\begin{figure*}[t]
\includegraphics[width=1\textwidth]{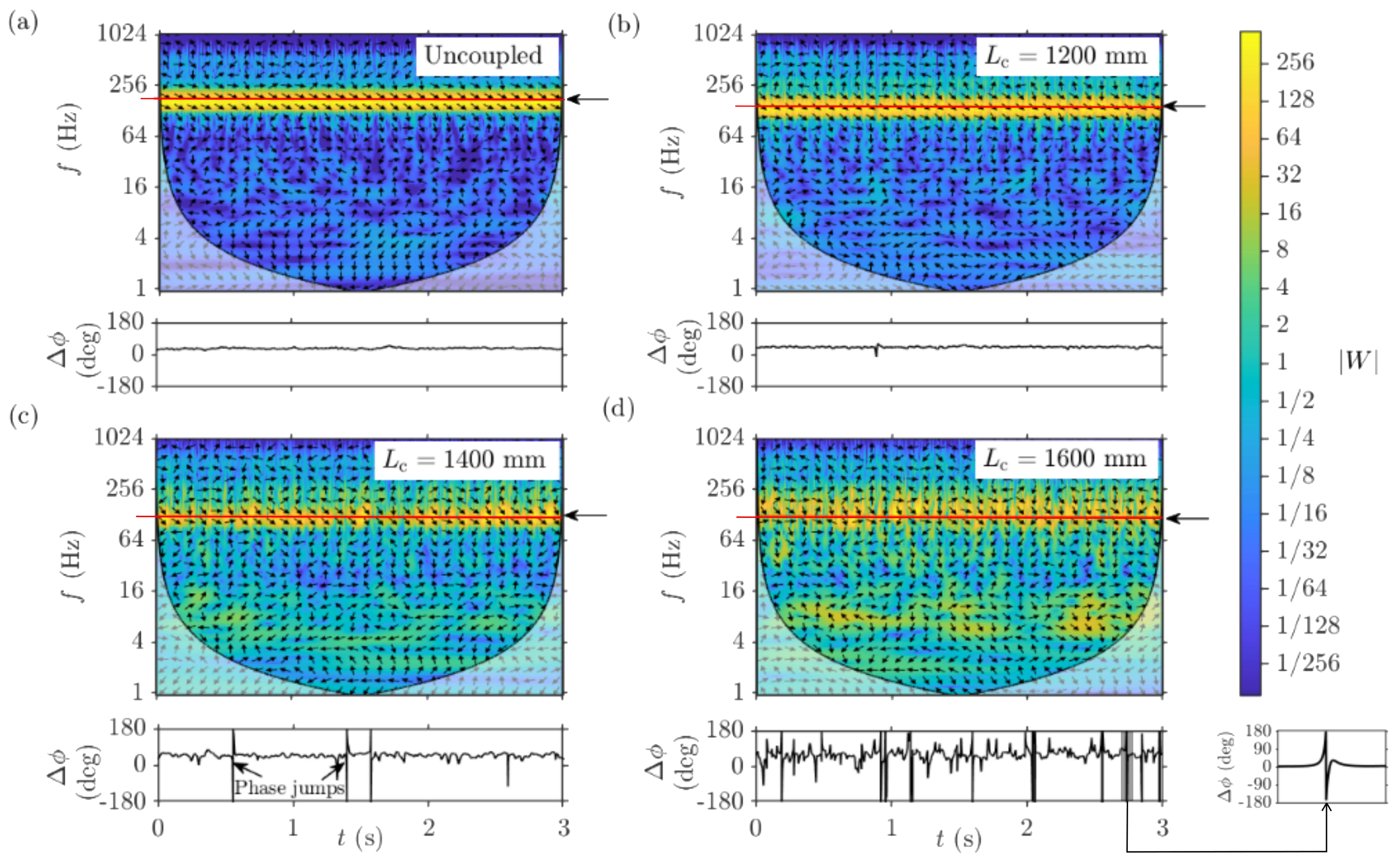}
\caption{\label{fig4} CWT plots between $p^\prime$ and $\dot{q}^\prime$ signals and the temporal variation of their phase difference $\Delta\phi$, calculated at the dominant frequency $f_d$ (indicated by a horizontal red line and an arrow in each CWT plot), for (a) the state of thermoacoustic instability in the absence of delayed acoustic self-feedback and (b-d) in the presence of delayed acoustic self-feedback in the combustor with increasing values of $L_{\text{c}}$. The inset in (d) shows a magnified view of a phase jump. The value of $d_{\text{c}}$ is kept constant at 25.4 mm.}
\end{figure*}

Since the amplitude of LCOs observed in the practical turbulent systems during the state of thermoacoustic instability could be different, we next examine the effectiveness of delayed acoustic self-feedback on the suppression of such instabilities of different amplitudes ($p^\prime_{0,\text{rms}}$) in the system (Fig. \ref{diff_amp_dia}a). The values of $L_{\text{c}} = 1400$ mm and $d_{\text{c}} = 25.4$ mm are fixed during these experiments, as they correspond to the regime of amplitude death of LCOs ($p^\prime_{0,\text{rms}} \approx 3200$ Pa) in Fig. \ref{fig2}. Here, the amplitude of LCOs before initiating the feedback is increased by increasing the fuel flow rate in the reactant mixture while keeping the equivalence ratio constant at the same time. In Fig. \ref{diff_amp_dia}(a),  we observe that LCOs of amplitudes $p^\prime_{0,\text{rms}} \leq 5000$ Pa can be completely suppressed with an optimum size of the coupling tube. In contrast, quenching of larger amplitude LCOs (i.e., $p^\prime_{0,\text{rms}} > 5000$ Pa) is not observed with this coupling tube. By "optimum size," we imply the dimensions (length and minimum internal diameter) of the connecting tube that are sufficient to achieve amplitude death in the system. The brown-colored data points on the right side of Fig. \ref{diff_amp_dia}(a) indicate no suppression of LCOs due to delayed acoustic self-feedback. Thus, we infer that a coupling tube with optimal values of $L_{\text{c}}$ and $d_{\text{c}}$ can mitigate thermoacoustic instability lesser than a certain critical value of $p^\prime_{0,\text{rms}}$.

Furthermore, we investigate the effects of the internal diameter of the coupling tube on the suppression of thermoacoustic instability (refer to Fig. \ref{diff_amp_dia}b). For two amplitudes of LCOs (i.e., $p^\prime_{0,\text{rms}} = 3400 \pm 150$ Pa and $4100 \pm 150$ Pa), we perform experiments with coupling tubes having five different internal diameters ranging from $d_{\text{c}} = 6.35 $ to $25.4$ mm, keeping the length of the coupling tube constant at $L_{\text{c}} = 1400$ mm. We observe that the occurrence of amplitude death depends on $d_{\text{c}}$, where a larger diameter coupling tube can quench such oscillations easily for the optimum value of $L_{\text{c}}$. Furthermore, LCOs with higher amplitude ($p^\prime_{0,\text{rms}} \approx 4100$ Pa) undergoes lower suppression than those with lower amplitudes ($p^\prime_{0,\text{rms}} \approx 3400$ Pa) for the same coupling parameters.

\subsection{\label{sec:Res_Disc_B} Temporal investigation of changes in the coupled acoustic pressure and global heat release rate oscillations}

As discussed in Sec.~\ref{sec:Introduction}, thermoacoustic instability occurs due to a closed-loop interaction between the global heat release rate ($\dot{q}^{\prime}$) and acoustic pressure ($p^{\prime}$) fluctuations. Several studies have recently shown that the onset of such instability in turbulent combustors is a consequence of mutual synchronization between these $p^{\prime}$ and global $\dot{q}^{\prime}$ fluctuations \cite{pawar2017thermoacoustic, godavarthi2018coupled, kasthuri2022coupled, murayama2019attenuation, guan2019chaos, dutta2019investigating, guan2022synchronization}. Therefore, quenching of thermoacoustic instability can be achieved by breaking the synchrony between these two oscillations. Thus, in this section, we investigate the effect of delayed acoustic self-feedback on the coupled behavior of the global $\dot{q}^{\prime}$ and $p^{\prime}$ fluctuations during the occurrence of amplitude death. We use different tools from the synchronization theory to study the coupled behavior of these oscillations. In Fig.~\ref{fig4}, we examine the locking of the instantaneous phases and frequencies of $p^{\prime}$ and global $\dot{q}^{\prime}$ signals using a cross-wavelet transform (CWT) \cite{grinsted2004application, pawar2019temporal}. In this plot, the regions of common spectral power for both signals are highlighted by the bright color, and the corresponding wrapped instantaneous relative phases between them are indicated by arrows. When the signals are synchronized, their CWT shows a larger magnitude of the spectral power throughout the time and a constant alignment of arrows at a particular phase difference. In contrast, the desynchronization of signals is indicated in the CWT by a near random distribution of the common spectral power and an arbitrary alignment of arrows in time.

\begin{figure}
\includegraphics[width=0.45\textwidth]{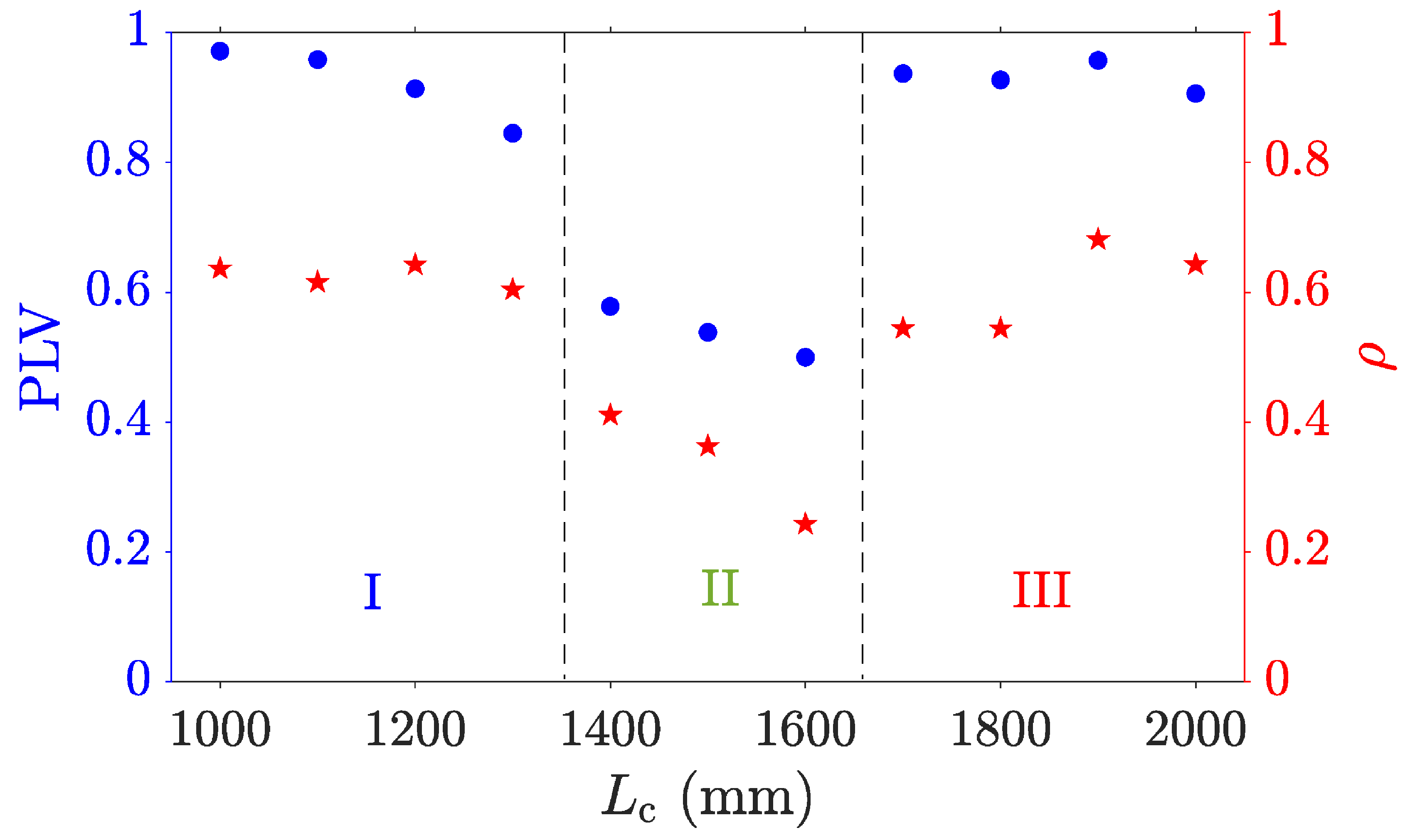}
\caption{\label{rho_plv} The variation of PLV and Pearson's correlation coefficient $\rho$ between the global heat release rate and acoustic pressure fluctuations under the influence of delayed acoustic self-feedback through connecting tubes of different lengths. The internal diameter of the coupling tube is kept constant at $d_{\text{c}} = 25.4$ mm, and $p^\prime_{0}$ is kept constant at 3400 $\pm$ 150 Pa.}
\end{figure}

In Fig. \ref{fig4}(a), we notice that, in the absence of delayed acoustic self-feedback in the system, a common frequency band occurs around the dominant frequency of 163.9 Hz in the CWT for the entire duration of the $p^{\prime}$ and global $\dot{q}^{\prime}$ signals, and the relative phase ($\Delta\phi$) between them is observed to stay constant around a mean value of $35.3^\circ$. Here, the instantaneous phase of the signals is calculated using the analytic signal method with the Hilbert transform \citep{pikovsky2003synchronization}. These behaviors of the frequency and phase indicate the existence of phase synchronization between the global heat release rate and acoustic pressure oscillations \cite{pawar2019temporal, kasthuri2022coupled}. When the system is subjected to delayed acoustic self-feedback for lower values of $L_{\text{c}}$ (see for $L_{\text{c}} = 1200$ mm in Fig.~\ref{fig4}b), i.e., in regime I of partial suppression of Fig. \ref{fig2}, we observe similar properties of phase synchronization between $p^{\prime}$ and global $\dot{q}^{\prime}$ signals as that witnessed in the absence of delayed acoustic self-feedback (Fig.~\ref{fig4}a). Just prior to the amplitude death of thermoacoustic instability (i.e., for $L_{\text{c}} = 1400$ mm in Fig. \ref{fig4}c), we notice the presence of discontinuities in the common frequency bands of CWT of $p^{\prime}$ and global $\dot{q}^{\prime}$ signals. These discontinuities can be easily seen in the plot of $\Delta\phi$, where we notice the number of phase jumps about a mean phase difference of these signals. During each phase jump, the value of $\Delta\phi$ shows a sudden change of $2\pi$ rad, and the phase jump happens due to the desynchronization of the signals \cite{pikovsky2003synchronization}. Furthermore, we find that these instances of desynchronized oscillations coincide with low-amplitude aperiodic oscillations observed in $p^{\prime}$ and global $\dot{q}^{\prime}$ signals (Fig.~\ref{fig3}d). Thus, in this condition of delayed acoustic self-feedback, the $p^{\prime}$ and global $\dot{q}^{\prime}$  exhibit a property of intermittent phase synchronization. 

During the state of amplitude death of thermoacoustic instability (i.e., for $L_{\text{c}} = 1600$ mm in Fig. \ref{fig4}d), we notice an increase in the discontinuities in the common frequency bands of CWT of $p^{\prime}$ and global $\dot{q}^{\prime}$ signals as compared to that seen in Fig. \ref{fig4}(c). This can also be confirmed from the $\Delta\phi$ plot, where we notice an increase in the number of phase jumps about a mean phase difference, which indicates desynchronization of $p^{\prime}$ and global $\dot{q}^{\prime}$ signals in the system.

In addition to CWT, we also use two other measures, called phase locking value (PLV) and Pearson's correlation coefficient ($\rho$), to detect the synchronization behavior of $p^{\prime}$ and global $\dot{q}^{\prime}$ signals for different values of $L_{\text{c}}$ shown in Fig. \ref{fig2}. PLV helps us to detect phase synchronization between two signals and is calculated as:
\begin{equation}
\text{PLV} = \dfrac{1}{N} \Bigg{\lvert} \sum\limits_{n=1}^N \text{exp}(i \Delta \phi) \Bigg{\lvert},
\end{equation}
where $N$ is the number of data points in the two signals, and $\Delta \phi$ is the instantaneous phase difference between the $p^{\prime}$ and global $\dot{q}^{\prime}$ signals. On the other hand, Pearson's correlation coefficient $\rho$ aids in finding the amplitude correlation between the signals \cite{gonzalez2002amplitude}, and is calculated as:
\begin{equation}
\rho = \dfrac{\text{Cov}(p^{\prime},\dot{q}^{\prime})}{\sigma_{p^{\prime}} \sigma_{\dot{q}^{\prime}}},
\end{equation}
where $\text{Cov}$ represents the covariance between two signals, and $\sigma$ denotes the standard deviation of a signal. We present the values of these measures in Fig. \ref{rho_plv}. We find that in the regime of partial suppression of thermoacoustic instability, both $p^{\prime}$ and global $\dot{q}^{\prime}$ signals are phase synchronized, confirmed from the PLV near 1. The presence of a high correlation between $p^{\prime}$ and global $\dot{q}^{\prime}$ signals in the regimes of oscillatory states is confirmed by the high positive value (between 0.55 and 0.7) of $\rho$. The value of $\rho$ drops to 0.25 in the regime of suppression of thermoacoustic instability (region II in Fig. \ref{rho_plv}), which happens due to the lack of correlation between desynchronized $p^{\prime}$ and global $\dot{q}^{\prime}$ signals \cite{pawar2017thermoacoustic}. 

\begin{figure*}
\includegraphics[width=1\textwidth]{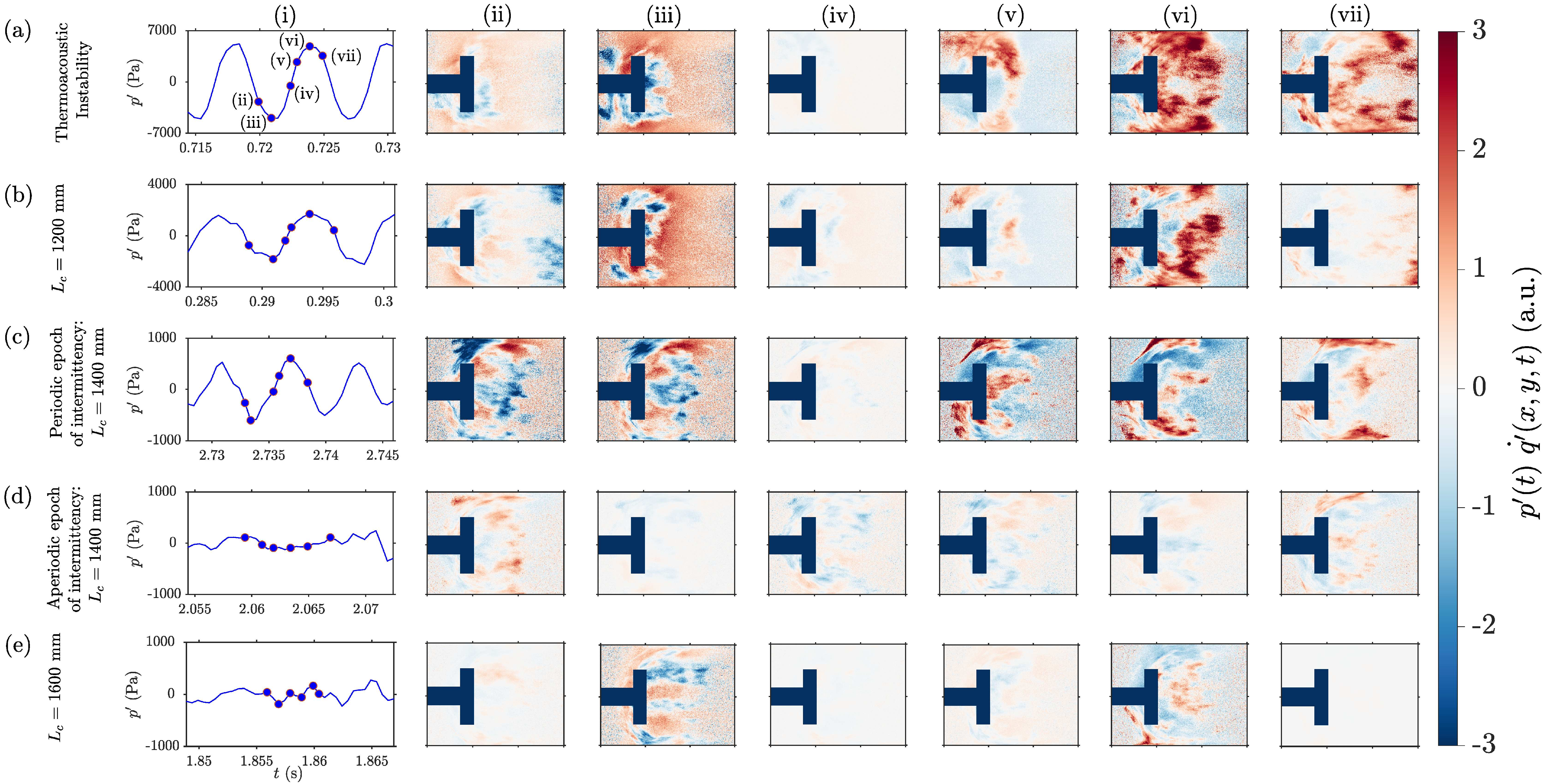}
\caption{\label{inst_pq} (i) Time instants of acoustic pressure signal, and (ii - vii) the corresponding instantaneous spatial distribution of local Rayleigh Index $p^\prime(t) \dot{q}^\prime (x,y,t)$ for (a) the state of thermoacoustic instability in the absence of delayed acoustic self-feedback, (b) the state of partial suppression of thermoacoustic instability corresponding to delayed acoustic self-feedback for $L_{\text{c}}=$ 1200 mm, (c) the periodic epoch and, (d) aperiodic epoch of intermittency ($L_{\text{c}}=$ 1400 mm), and (e) the state of amplitude death for $L_{\text{c}}=$ 1600 mm. The internal diameter of the coupling tube is kept constant at $d_{\text{c}}=$ 25.4 mm, and $p^\prime_{0,\text{rms}}$ is kept constant at around 3400 Pa.}
\end{figure*}

Thus, we observe that the extent of synchronization between the coupled $p^{\prime}$ and global $\dot{q}^{\prime}$ signals decreases during the mitigation of thermoacoustic instability in a turbulent combustor, where the signals change their behavior from the state of phase synchronized periodicity to desynchronized aperiodicity via intermittent synchronized oscillations. Next, we look at the effectiveness of delayed acoustic self-feedback to disrupt the complex closed-loop interactions between hydrodynamic, acoustic, and heat release rate fluctuations that exist inside the turbulent combustor during the state of thermoacoustic instability.

\subsection{\label{sec:Res_Disc_C} Spatiotemporal analysis of acoustic power production in the turbulent flow field during different states of self-delayed feedback}

In this section, we compare the spatiotemporal changes in the reaction field of the combustor as thermoacoustic instability is suppressed through delayed acoustic self-feedback. The local heat release rate fluctuations $\dot{q}^{\prime}(x,y,t)$ in the combustor are obtained by the high-speed CH* chemiluminescence imaging on the flame. In Fig. \ref{inst_pq}, we examine the local acoustic power $p^\prime (t) \dot{q}^{\prime}(x,y,t)$ field (see panels ii-vii) at some representative time instances marked on $p^{\prime}$ signal (see panel i). Regions with $p^\prime (t) \dot{q}^{\prime}(x,y,t)$ $>$ $0$ represent acoustic power sources, and regions with $p^\prime (t) \dot{q}^{\prime}(x,y,t)$ $<$ $0$ represent acoustic power sinks \cite{krishnan2019emergence}. Due to the periodic formation of large-scale vortical structures at the dump plane and the tip of the bluff body, we observe the coherent production of acoustic power sources over large clusters in the reaction field during thermoacoustic instability (Fig. \ref{inst_pq}a). In these coherent regions of acoustic power, the magnitude of the acoustic power sources is significantly greater than their negative counterparts in regions of acoustic power sink. The regions of acoustic power sources grow in size when the acoustic pressure fluctuations approach local extrema (Figs. \ref{inst_pq}a-iii and  \ref{inst_pq}a-vii), and they decrease in their spatial size when the acoustic pressure fluctuations $p^\prime$ approach the mean value of zero (Figs. \ref{inst_pq}a-ii and  \ref{inst_pq}a-iv).

A similar trend of change in the size of acoustic power sources over time is observed when the turbulent combustor is subjected to delayed acoustic self-feedback by coupling it with a connecting tube of length $L_{\text{c}} = 1200$ mm (Figs. \ref{inst_pq}b-ii and \ref{inst_pq}b-vii)), and also during a periodic epoch of intermittency observed in Fig. \ref{inst_pq}(c) for $L_{\text{c}} = 1400$ mm. High spatial coherence of acoustic power sources is observed at a large scale in Fig. \ref{inst_pq}(b)-(vi) downstream of the bluff body. However, the magnitude and spatial coherence of acoustic power sources observed in Fig. \ref{inst_pq}(b) and Fig. \ref{inst_pq}(c) are low compared to that observed during the state of thermoacoustic instability. This is accompanied with an increase in the suppression of thermoacoustic instability from Fig. \ref{inst_pq}(b) to Fig. \ref{inst_pq}(c). During the aperiodic epoch of intermittency, we do not observe any large coherent regions of acoustic power in Fig. \ref{inst_pq}(d), similar to that observed during the periodic epochs of intermittency. However, we observe small scale patches of acoustic power sources downstream (Figs. \ref{inst_pq}d-i and  \ref{inst_pq}d-vii) and upstream (Fig. \ref{inst_pq}d-iv) of the bluff body. We believe that these small scale acoustic power sources result from small scale vortices shed from the dump plane and the tip of the bluff body that do not grow in size and combust shortly after they are formed.

Finally, when the length of the connecting tube is increased to $L_{\text{c}} = 1600$ mm, we notice the occurrence of amplitude death of thermoacoustic instability, where the acoustic pressure exhibits low-amplitude chaotic oscillations as seen in Fig. \ref{inst_pq}(e)-(i). During this state, the spatial distribution of $p^\prime (t) \dot{q}^{\prime}(x,y,t)$ seems granular and devoid of a discernible pattern for different instants of time. These grainy structures can be considered as disordered patterns (or incoherent) patterns in the spatial dynamics of the turbulent reacting field. We hypothesize that the presence of several small scale vortices in the flow field results in incoherent acoustic power production. As discussed in Sec. \ref{sec:Introduction}, the occurrence of thermoacoustic instability in a turbulent combustion system is a complex process and happens due to nonlinear interaction between the flame, the flow, and the acoustic field of a combustor. The application of optimal delayed acoustic self-feedback disrupts this complex interaction in the turbulent combustor, thus mitigating thermoacoustic instability.

\section{\label{sec:Conclusion} Conclusions}

In summary, we demonstrate the mitigation of thermoacoustic instability (i.e., limit cycle oscillations) in a bluff body stabilized turbulent combustor by inducing a delayed acoustic self-feedback in the system. The complex closed-loop coupling between acoustics, hydrodynamics, and flame dynamics developed during thermoacoustic instability is disrupted when optimal delayed acoustic self-feedback is introduced in the system. This delayed self-feedback is achieved by coupling the acoustic field of the combustor to itself, using a connecting tube attached near the pressure anti-node position of the acoustic standing wave. We observe a significant suppression of thermoacoustic instability when the length of the coupling tube is approximately 3/8 times the wavelength of the fundamental acoustic mode established during the state of thermoacoustic instability in the combustor. The amplitude of the acoustic pressure fluctuations during the state of suppression is comparable to that observed for the state of stable operation (combustion noise). As the length of the coupling tube is increased, we find that the mitigation of thermoacoustic instability is associated with a gradual decrease in the amplitude of acoustic pressure oscillations and a shift in their dominant frequency toward lower values. Furthermore, we notice that the dynamical behavior of acoustic pressure fluctuations changes from the state of limit cycle oscillations to low-amplitude chaotic oscillations via intermittent oscillations during the suppression of thermoacoustic instability. The synchronization of the global heat release rate and acoustic pressure fluctuations, observed during the uncoupled state of thermoacoustic instability, breaks down gradually and the oscillations become desynchronized as the system approaches the state of amplitude death. During this state, we do not observe any coherent structures of acoustic power production in the reaction field as witnessed during thermoacoustic instability. We also notice the disintegration of acoustic power sources in the flow field during the state of suppression of oscillations, which happens due to the destruction of the local coupling between acoustic pressure and heat release rate fluctuations in the spatial field of the combustor. 

A similar methodology based on self-acoustic feedback has been recently used by Srikanth \textit{et al.} \cite{srikanth2021selfcoupling} to mitigate thermoacoustic instabilities in a horizontal laminar Rijke tube. The Rijke tube consists of an electrically heated wire mesh as a compact heat source in the laminar flow field. We observed many similarities in the behavior of self-coupled laminar and turbulent thermoacoustic systems, which include an optimum length of the coupling tube needed to quench thermoacoustic instability, the suppression of thermoacoustic instability being limited by the amplitude of these oscillations, and the variation of frequency of acoustic pressure oscillations due to the change in the coupling tube length. Nevertheless, due to the presence of a laminar flow field in the Rijke tube system, the suppression of thermoacoustic instability happens abruptly from the
state of limit cycle oscillations to a steady state when the length of the coupling tube is greater than a critical value. However, in turbulent combustors, we observe that the suppression of thermoacoustic instability is nearly gradual, where the state of intermittency is noticed prior to amplitude death of thermoacoustic instability. In the amplitude death state observed in turbulent combustors,
the pressure fluctuations are chaotic and show significantly larger
amplitude than zero.

Thus, we show that delayed acoustic self-feedback achieved through a single connecting tube provides a promising control mechanism to suppress thermoacoustic instability in turbulent combustors. Unlike in traditional active closed-loop controls used in thermoacoustic studies, we do not need to preprocess the acoustic pressure signal acquired from the combustor prior to feedback using any electromechanical devices in the method of delayed acoustic self-feedback. Therefore, we believe this methodology based on delayed acoustic self-feedback opens up novel, cost-effective ways to mitigate thermoacoustic instability in turbulent combustion systems used in practical gas turbine engines.

\begin{acknowledgments}
We thank Mr. S. Thilagaraj and Mr. Anand for their help in conducting the experiments. The authors also thank Manikandan Raghunathan for fruitful discussions regarding the spatiotemporal analysis. This work is supported by the J. C. Bose Fellowship (No. JCB/2018/000034/ SSC) and the IoE initiative (SB/2021/0845/AE/MHRD/002696) from the Department of Science and Technology, Government of India.
\end{acknowledgments}

\section*{Data Availability Statement}
The data that support the findings of this study are available from the corresponding author upon reasonable request.

\bibliography{aipsamp}

\end{document}